\documentclass[final,5p,times,onecolumn]{elsarticle}

\usepackage{graphicx}
\usepackage{amsmath,amssymb,amsfonts}
\usepackage{xcolor}
\usepackage{lineno}

\usepackage[colorlinks=true,citecolor=blue,urlcolor=blue]{hyperref}

\def\deg{^\circ}
\def\gtorder{\mathrel{\raise.3ex\hbox{$>$}\mkern-14mu
 \lower0.6ex\hbox{$\sim$}}}
\def\ltorder{\mathrel{\raise.3ex\hbox{$<$}\mkern-14mu
 \lower0.6ex\hbox{$\sim$}}}

\begin{document}

\begin{frontmatter}

\title{\texorpdfstring{Measuring short-range correlations and quasi-elastic cross sections in A(e,e') at $x>1$ and modest Q$^2$}{Measuring short-range correlations and quasi-elastic cross sections in A(e,e') at x>1 and modest Q2}}

\author[Tsinghua]{Y.P. Zhang}
\author[Tsinghua,Virginia]{Z.H. Ye}
\author[Virginia]{D. Nguyen}
\author[Paris]{P. Aguilera}
\author[Syracuse]{Z. Ahmed}
\author[TAMUK]{H. Albataineh}
\author[JLAB]{K. Allada}
\author[KENT]{B. Anderson}
\author[Halifax]{D. Anez}
\author[CALIF]{K.~Aniol}
\author[Glasgow]{J. Annand}
\author[LBL,Argonne]{J. Arrington}
\author[William]{T. Averett}
\author[Virginia]{H. Baghdasaryan}
\author[China]{X. Bai}
\author[NRCN]{A. Beck}
\author[NRCN]{S. Beck}
\author[Catania]{V.~Bellini}
\author[Dequense]{F.~Benmokhtar}
\author[JLAB]{A. Camsonne}
\author[Hampton]{C. Chen}
\author[JLAB]{J.-P. Chen}
\author[Virginia]{K. Chirapatpimol}
\author[INFN]{E.~Cisbani}
\author[JLAB]{S.~Covig~Dusa}
\author[JLAB,Ohio]{M.~M.~Dalton}
\author[Virginia]{D.~Day}
\author[MIT]{W.~Deconinck}
\author[Ecole]{M.~Defurne}
\author[Temple]{D.~Flay}
\author[Tennessee]{N.~Fomin}
\author[Pittsburgh]{M.~Friend}
\author[Temple]{E.~Fuchey}
\author[INFN]{F.~Garibaldi}
\author[JLAB]{D.~Gaskell}
\author[MIT]{S.~Gilad}
\author[Rutgers]{R.~Gilman}
\author[Kharkov]{S.~Glamazdin}
\author[Virginia]{C.~Gu}
\author[Hampton]{P.~Gu\`eye}
\author[Virginia]{C.~Hanretty}
\author[JLAB]{J.-O.~Hansen}
\author[Virginia]{M.~Hashemi~Shabestari}
\author[JLAB]{D.~W.~Higinbotham}
\author[Duke]{M.~Huang}
\author[CALIF]{S.~Iqbal}
\author[Virginia]{G.~Jin}
\author[Virginia]{N.~Kalantarians}
\author[Seoul]{H.~Kang}
\author[MIT]{A.~Kelleher}
\author[TLV]{I.~Korover}
\author[JLAB]{J.~LeRose}
\author[Indiana]{J.~Leckey}
\author[Hampshire]{S.~Li}
\author[Virginia]{R.~Lindgren}
\author[KENT]{E.~Long}
\author[Blacksburg]{J.~Mammei}
\author[Florida]{P. Markowitz}
\author[JLAB]{D. Meekins}
\author[JLAB]{R.~Michaels}
\author[Jozef]{M.~Mihovilovi\v{c}}
\author[MIT]{N.~Muangma}
\author[France]{C.~Munoz~Camacho}
\author[Virginia]{B.~E.~Norum}
\author[Mississippi]{Nuruzzaman}
\author[MIT]{K.~Pan}
\author[Hampshire]{S.~Phillips}
\author[TLV]{E.~Piasetzky}
\author[TLV,Austin]{I.~Pomerantz}
\author[Temple]{M.~Posik}
\author[Norfolk]{V.~Punjabi}
\author[Duke]{X.~Qian}
\author[JLAB]{Y.~Qiang}
\author[Lanzhou]{X.~Qiu}
\author[Argonne]{P.~E.~Reimer}
\author[Syracuse]{A.~Rakhman}
\author[Virginia,Massachusetts]{S.~Riordan}
\author[Hebrew]{G.~Ron}
\author[Virginia]{O.~Rondon-Aramayo}
\author[KENT]{L.~Selvy}
\author[Yerevan]{A.~Shahinyan}
\author[TLV]{R.~Shneor}
\author[Ljubljana,Jozef]{S.~\v{S}irca}
\author[Hampshire]{K.~Slifer}
\author[Temple]{N.~Sparveris}
\author[Virginia]{R.~Subedi}
\author[MIT]{V.~Sulkosky}
\author[Virginia]{D.~Wang}
\author[KENT]{J.~W.~Watson}
\author[DOMINION]{L.~B.~Weinstein}
\author[JLAB]{B.~Wojtsekhowski}
\author[JLAB]{S.~A.~Wood}
\author[TLV]{I.~Yaron}
\author[JLAB]{J.~Zhang}
\author[Rutgers]{Y.~W.~Zhang}
\author[William]{B.~Zhao}
\author[Virginia]{X.~Zheng}
\author[Hefei]{P.~Zhu}
\author[Hampshire]{R.~Zielinski}

\address[Tsinghua]{Department of Physics, Tsinghua University, Beijing, China}
\address[Virginia]{University of Virginia, Charlottesville, Virginia 22904}
\address[Paris]{Institut de Physique Nucleaire (UMR 8608), CNRS/IN2P3 - Universite Paris-Sud, France}
\address[Syracuse]{Syracuse University, Syracuse, New York 13244}
\address[TAMUK]{Texas A\&M University - Kingsville, Kingsville, Texas, 78363}
\address[JLAB]{Thomas Jefferson National Accelerator Facility, Newport News, Virginia 23606}
\address[KENT]{Kent State University, Kent, Ohio 44242}
\address[Halifax]{Saint Mary's University, Halifax, Nova Scotia, Canada}
\address[CALIF]{California State University, Los Angeles, CA 90032}
\address[Glasgow]{University of Glasgow, Glasgow, Scotland, UK}
\address[LBL]{Lawrence Berkeley National Laboratory, Berkeley, California}
\address[Argonne]{Physics Division, Argonne National Laboratory, Argonne, Illinois}
\address[William]{College of William \& Mary, Williamsburg, Virginia}
\address[China]{China Institute of Atomic Energy, Beijing, China}
\address[NRCN]{Nuclear Research Center Negev, Beer-Sheva, Israel}
\address[Catania]{Universita di Catania, Italy}
\address[Dequense]{Duquesne University, Pittsburgh, PA}
\address[Hampton]{Hampton University, Hampton, Virginia}
\address[INFN]{INFN, Sezione Sanità and Istituto Superiore di Sanità, Rome, Italy}
\address[Ohio]{Ohio University, Athens, Ohio}
\address[MIT]{Massachusetts Institute of Technology, Cambridge, MA}
\address[Ecole]{CEA Saclay, Gif-sur-Yvette, France}
\address[Temple]{Temple University, Philadelphia, Pennsylvania}
\address[Tennessee]{University of Tennessee, Knoxville, TN}
\address[Pittsburgh]{Carnegie Mellon University, Pittsburgh, PA}
\address[Rutgers]{Rutgers University, Piscataway, New Jersey}
\address[Kharkov]{Kharkov Institute of Physics and Technology, Ukraine}
\address[Duke]{Duke University, Durham, North Carolina}
\address[Seoul]{Seoul National University, Korea}
\address[TLV]{Tel Aviv University, Israel}
\address[Indiana]{Indiana University, Bloomington, Indiana}
\address[Hampshire]{University of New Hampshire, Durham, NH}
\address[Blacksburg]{Virginia Tech, Blacksburg, Virginia}
\address[Florida]{Florida International University, Miami, Florida}
\address[Jozef]{Jozef Stefan Institute, Ljubljana, Slovenia}
\address[France]{Université Blaise Pascal/IN2P3, France}
\address[Mississippi]{Mississippi State University, Starkville, MS}
\address[Austin]{The University of Texas at Austin, Austin, Texas}
\address[Norfolk]{Norfolk State University, Norfolk, Virginia}
\address[Lanzhou]{Lanzhou University, China}
\address[Massachusetts]{University of Massachusetts, Amherst, Massachusetts}
\address[Hebrew]{Hebrew University of Jerusalem, Israel}
\address[Yerevan]{Yerevan Physics Institute, Armenia}
\address[Ljubljana]{Faculty of Mathematics and Physics, University of Ljubljana, Slovenia}
\address[DOMINION]{Old Dominion University, Norfolk, Virginia}
\address[Hefei]{University of Science and Technology, Hefei, China}
  
\end{frontmatter}

\twocolumn[
  \begin{@twocolumnfalse}
   \begin{center}
       \section*{Abstract}
   \end{center} 
\hspace{0.5cm} We present results from the Jefferson Lab E08-014 experiment, investigating short-range correlations (SRC) through measurements of inclusive quasi-elastic scattering from $^2$H, $^3$He, $^4$He, $^{12}$C, $^{40}$Ca, and $^{48}$Ca. The kinematics were selected to isolate scattering from SRCs, yielding a plateau in the A/$^2$H cross-section ratios due to the universal two-body structure of the 2N-SRCs in light and heavy nuclei. We observe approximate plateaus in the A/$^2$H ratios and provide the first extractions of the A/$^2$H ratio for $^{40}$Ca and $^{48}$Ca. We also examine the A/$^3$He ratio, aiming to identify three-nucleon SRCs (3N-SRCs). Following the approach for isolating 2N-SRCs, searching for 3N-SRC dominance involved measuring the A/$^3$He cross section ratio at modest-to-large $Q^2$ values and looking for a plateau ratios for $x \gtorder 2.5$. This was not observed in the data, and in fact increasing $Q^2$ values moved the data \textit{further away} from the predicted plateau. We show here that, when analyzed in terms of the struck nucleon's light-cone momentum, the data exhibited the expected trend, progressively approaching the predicted 3N-SRC plateau. These observations suggest that future measurements at higher energies may facilitate a definitive isolation and identification of 3N-SRCs.

\date{\today}

\vspace{1cm}
\end{@twocolumnfalse}
]

\section{Introduction}
Short-range correlations (SRCs) are transitory configurations of nucleons with large relative momenta and modest total momentum. They are dominated by two-nucleon SRCs (2N-SRC) but may have significant contributions from three-nucleon SRCs (3N-SRC) in specific kinematic regimes. SRCs do not arise from  mean-field potentials derived from models such as the shell model. Instead, they arise from ephemeral fluctuations in the nuclear ground state, driven by the strong, short-range component of the nucleon-nucleon (NN) interaction, which includes both an attractive tensor force and a deep repulsive core at sub-fm distances. These short-range components are pervasive and consequently all nuclei display SRCs with universal characteristics. They contribute to the high-momentum tails observed in nucleon momentum distributions and the depletion of spectral strength seen in exclusive electron scattering experiments~\cite{Lapikas:1999ss, Sargsian:2002wc, Arrington:2012ax, Dalal:2022zkg}. A comprehensive study of SRCs is essential for a deeper understanding of nuclear forces and the detailed structure of atomic nuclei. Significant progress has been made in both theoretical and experimental investigations in recent years~\cite{Sargsian:2002wc, Frankfurt:1988nt, Arrington:2011xs, CiofidegliAtti:2015lcu, Hen:2016kwk, Fomin:2017ydn, Arrington:2022sov}.

SRCs have been studied extensively through quasi-elastic (QE) electron scattering, where high-energy electron beams knock high-momentum nucleons out of nuclei at large momentum transfer~\cite{Sargsian:2002wc, Arrington:2022sov}. Inclusive electron scattering experiments, A(e,e'), which measure only the scattered electrons, have been conducted at SLAC~\cite{Frankfurt:1993sp} and Jefferson Lab (JLab)~\cite{Arrington:1998ps, CLAS:2003eih, CLAS:2005ola, Fomin:2011ng, Ye:2017ivm, CLAS:2019vsb, Nguyen:2020wrr, Li:2022fhh, Li:2024rzf}, confirming the basic predictions of the SRC model~\cite{Frankfurt:1988nt, Frankfurt:1993sp} and quantifying the relative contribution of SRCs in nuclei~\cite{Fomin:2017ydn, Arrington:2022sov}. Exclusive measurements of QE reactions, which reconstruct back-to-back high-momentum nucleons from the breakup of 2N-SRC pairs in addition to detecting the scattered electron, have demonstrated a dominance of proton-neutron (pn) pairs over proton-proton (pp) or neutron-neutron (nn) pairs~\cite{Tang:2002ww, Shneor:2007lly, Subedi:2008zz, Korover:2014wqo, CLAS:2018xvc}. This dominance is a manifestation of the strong tensor force, which preferentially binds pn pairs, making them more likely to form SRCs~\cite{Schiavilla:2006xx, Alvioli:2007zz, CiofidegliAtti:2015lcu, CLAS:2020mom}. The observed pn dominance was confirmed in inclusive studies of nuclei with similar mass but different neutron-to-proton ratios, such as $^{48}$Ca/$^{40}$Ca~\cite{Nguyen:2020wrr} and $^3$He/$^3$H~\cite{Li:2022fhh}. While this isospin dependence is a universal feature in heavier nuclei, recent results suggest a weaker preference for pn pairs in light nuclei~\cite{Li:2022fhh}.

The first evidence for 2N-SRCs came from  inclusive QE per-nucleon cross-section ratios of nuclei to deuterium, i.e. $R=\frac{\sigma_A}{\sigma_{^2H}}\frac{2}{A}$ (referred to as simply A/$^2$H ratios in this paper), which revealed a distinct plateau in the region $1.4 < x < 2.0$~\cite{Frankfurt:1993sp} at $Q^2 \gtorder 1.4$~GeV$^2$. 
Here, $Q^2$ is the four-momentum transfer squared, and $x$ is the Bjorken variable, both of which are determined from the beam energy ($E_0$) and the measured scattered electron momentum ($E^\prime$) and angle ($\theta_e$). In this framework, $Q^2 = 4E_0E^\prime \sin^2(\theta_e/2)$ and $x = Q^2 / (2m_p \nu)$, where $m_p$ is the proton mass and $\nu = E_0 - E^\prime$. Following this pioneering study, extensive measurements at JLab~\cite{Fomin:2011ng, CLAS:2019vsb, Li:2022fhh, Li:2024rzf} revealed the predicted 2N-SRC dominance~\cite{Frankfurt:1988nt, Frankfurt:1993sp} and verified the kinematic conditions for which 2N-SRCs dominate over mean-field contributions. The value of the A/$^2$H ratio in the plateau region, denoted $a_2$(A), reflects the relative contribution of SRCs in heavier nuclei compared to the deuteron. However, the motion of SRCs in nuclei with $A > 2$ enhances the cross-section ratio by $\sim$10–20\% above what is predicted by simply counting SRCs~\cite{Fomin:2011ng, Weiss:2020bkp}. This enhancement increases with nuclear mass (A), indicating a greater prevalence of nucleons in SRC pairs in heavier nuclei due to increased nucleon-nucleon interactions in denser nuclear environments, with the trend saturating at A~$\geq 12$~\cite{Arrington:2012ax}.

The A/$^3$He per-nucleon cross-section ratios (i.e. $R=\frac{\sigma_A}{\sigma_{^3He}}\frac{3}{A}$) also exhibit the predicted 2N-SRC plateau~\cite{CLAS:2003eih, CLAS:2005ola, Fomin:2011ng, Ye:2017ivm}, but the SRC model also predicts a second plateau corresponding to the onset of 3N-SRCs above $x=2$. For 2N-SRCs, the model predicts that SRCs dominate for momenta above the typical mean-field momenta in nuclei, while for 3N-SRCs, it is less clear when 2N-SRC contributions are small enough that the 3N-SRC contributions can be isolated. The predicted 3N-SRC plateau appeared to be observed in the $^4$He/$^3$He ratio at $x \geq 2.25$ and $Q^2 \sim 1.6$~GeV$^2$ by CLAS~\cite{CLAS:2005ola}, while a Hall C experiment at $Q^2 \sim 2.7$~GeV$^2$ reported a similar trend, albeit with large statistical uncertainties~\cite{Fomin:2011ng, Day:2018nja}. However, the hypothesis that 3N-SRCs dominated at $Q^2\approx1.6$~GeV$^2$ and large $x$ was ruled out by the initial publication of the present results~\cite{Ye:2017ivm}, with the limited momentum resolution of the CLAS spectrometer being the apparent cause of the observed plateau~\cite{Higinbotham:2009zz}. It has been argued that further searches for the onset of 3N-SRCs should focus on high-statistics measurements at large $Q^2$ values ($Q^2 \gtorder 3$~GeV$^2$)~\cite{Fomin:2017ydn, Day:2018nja, Sargsian:2019joj, Fomin:2023gdz}. It has also been suggested that focusing on very light nuclei, in particular $^3$H and $^3$He, may provide easier access to information on 3N-SRCs~\cite{Li:2024rzf}. At this time, the existence of 3N-SRCs remains an open question that warrants further investigation~\cite{Ye:2018jth, Arrington:2022sov, Fomin:2023gdz} and only limited information on their structure exists~\cite{CLAS:2010yvl, Li:2024rzf}.

\section{Experimental Overview and Data Analysis}

This paper reports the results from the JLab E08-014 experiment, which measured scattered electrons and extracted inclusive QE cross sections for $1 \ltorder x < 3.0$ and moderate $Q^2$ values~\cite{Ye_Thesis}. The previously published $^4$He/$^3$He ratio revealed the absence of the predicted 3N-SRC signal~\cite{Ye:2017ivm}, while the $^{48}$Ca/$^{40}$Ca ratio provided the first examination of the isospin effect in 2N-SRCs using the inclusive scattering~\cite{Nguyen:2020wrr}. Here, we present the absolute cross sections for all nuclei measured in this experiment, including $^2$H and $^{12}$C targets in addition to $^4$He, $^3$He, $^{48}$Ca, and $^{40}$Ca. We also present the A/$^2$H and A/$^3$He ratios to explore 2N-SRCs and 3N-SRCs. 

JLab experiment E08-014~\cite{E08-014} utilized a 3.356~GeV unpolarized electron beam with currents ranging from 40 to 120 $\mu$A, provided by the Continuous Electron Beam Accelerator Facility (CEBAF). The beam was directed onto several cryogenic targets, including $^2$H, $^3$He, and $^4$He, which were housed in 20~cm aluminum cells. These cells were mounted on a vertically adjustable ladder within a vacuum chamber. Solid foil targets, including $^{12}$C, $^{40}$Ca, and $^{48}$Ca, were positioned below the cryogenic targets. The entire target ladder was cooled to 20 K using a cryotarget cooling system~\cite{Alcorn:2004sb}. 

Two high-resolution magnetic spectrometers (HRS-L and HRS-R)~\cite{Alcorn:2004sb} were used to measure the quasi-elastically scattered electrons. Each spectrometer consists of two quadrupoles (Q1 and Q2), a dipole (D), and a final quadrupole (Q3). The magnetic fields of these components were set to match the standard optics tune for the desired central momentum value~\cite{Alcorn:2004sb}. However, the Q3 magnet on HRS-R encountered a power supply issue, resulting in a 15\% reduction in the field. This necessitated the acquisition of new optics data and the recalibration of the optics matrices offline to accurately account for the mismatch in the Q3 field (see Chapter 4 in Ref.~\cite{Ye_Thesis}).

The detector setup in each HRS includes two vertical drift chambers (VDCs), two scintillator hodoscope planes (S1 and S2), a gas Cherenkov counter (GC), and two layers of electromagnetic calorimeters (ECAL). The VDCs track charged particles exiting from Q3, allowing the determination of their momenta and angles at the reaction points using the HRS magnet optics. The signals from S1, S2, and GC formed triggers for the events recorded by the data acquisition (DAQ) system, with scalers monitoring detection efficiencies and computer dead time. The GC and ECAL also serve as particle identification (PID) detectors to differentiate between electrons and pions.

Data were taken with the spectrometers positioned at angles between 21$^\circ$ and 28$^\circ$, with one or more momentum settings at each angle to cover a wide range of $x$. Detailed kinematic settings can be found in the supplemental materials~\cite{supplemental}.

\subsection{Electron Reconstruction and Selection}

The HRS spectrometers detected scattered electrons as well as other particles, primarily $\pi^{-}$ mesons. The Hall A standard offline data analysis reconstructed the scattered electron momentum ($P_e$), in-plane and out-of-plane angles ($\phi_e$, $\theta_e$), and vertex position ($Z_{react}$) using the VDC track information and optics reconstruction matrices~\cite{Alcorn:2004sb}. For the HRS-R, the reconstruction matrices from previous experiments were updated to correct for the reduced Q3 magnetic field~\cite{Ye_Thesis}.

To minimize uncertainties associated with the spectrometer acceptance, in particular for the long targets, strict acceptance cuts were applied: $|\delta P_e| \leq 4$\%, $|\delta \theta_e| \leq 30$ mrad, and $|\delta \phi_e| \leq 30$ mrad. For gas targets, an additional cut of $|\delta Z_{react}| \leq 7$~cm was implemented to exclude events originating from the target entrance and exit windows. Residual contamination was evaluated using data collected from measurements with two thick aluminum foils designed to replicate the target windows. These studies indicate negligible contamination mixing with genuine gas target events ($<$0.1\%) from window events due to reconstruction inaccuracies. It is worth noting that in Ref.~\cite{Nguyen:2020wrr}, only the short targets ($^{40}$Ca and $^{48}$Ca) were included, allowing for slightly wider acceptance cuts and smaller statistical uncertainties. Calibrated GC and ECAL signals were used to apply PID cuts, achieving over 99\% efficiency for identifying electrons, with negligible pion contamination (see Sec. 5.5.3 of Ref.~\cite{Ye_Thesis}).

The target cooling system served to reduce density fluctuations in the cryogenic targets and protect the foil targets from beam-induced heating. However, since only one end of the 20~cm extended target cells was mounted on the cooled ladder, non-uniform density fluctuations, often referred to as boiling effects, were observed across the cryogenic targets. A new procedure was performed to correct for the non-uniform boiling effect and obtain in-beam target densities for all three cryogenic targets (See Section 2 in the supplemental materials~\cite{supplemental} for details).

\subsection{Cross Section Extraction}
For each setting, we binned the data in $x$ bins and extracted the Born cross section by applying the ratio of the experimental to simulated yield as a correction to the model cross section used in the simulation:
\begin{equation}
    \sigma_{EX}(x_i, \theta_0) = {Y^i_{EX}}/ {Y^i_{MC}} \cdot \sigma_{model}(x_i, \theta_0),
    \label{xs_eq}
\end{equation}
where the experimental (model) differential cross section in the $i$th bin is given as $\sigma_{EX(model)}(x_i)$ and $x_i$ is the central value of the bin and $\theta_0$ is the central angle of the spectrometer. 

The experimental yield, $Y^i_{EX}$, is defined as: 
\begin{equation}
    Y^i_{EX} = \frac{N^i_{EX}} {N_e \eta_{tg} \epsilon_{eff}}, 
\end{equation}
where $N^i_{EX}$ is the number of scattered electrons in the $x$ bin after applying acceptance and PID cuts, and has also been corrected for computer dead time. $N_e = Q_e / e$ represents the total number of incoming electrons, derived from the accumulated beam charge used to acquire the data. The areal density of scattering centers is given as: $\eta_{tg} = \rho \cdot \Delta L_{tg} \cdot N_a / A$, where $\rho$ is the target density, $\Delta L_{tg}$ is the target length, $N_a$ is Avogadro’s number, and A is the atomic mass number. For cryogenic targets, $\rho \cdot \Delta L_{tg}$ has been corrected for the non-uniform boiling effect as described in Section 2 of the supplemental materials. The detection efficiency, $\epsilon_{eff}$, includes the efficiencies for triggering, tracking, and PID. The analysis shows that these efficiencies are consistently above 99\%, allowing us to set $\epsilon_{eff} = 1$ with a 1\% systematic uncertainty.

The Monte Carlo yield, $Y^i_{MC}$, is defined as:
\begin{equation}
    Y^i_{MC} = \frac{\Delta E'_{MC}\Delta \Omega_{MC}}{N^{gen}_{MC}} \cdot \sum_{j\in i} \sigma^{rad}_{model}(x_j), 
\end{equation}
where $\Delta E’_{MC}$ and $\Delta \Omega_{MC}$ are the ranges of scattered electron energies and solid angles defined by the same acceptance cuts applied to the real data, and $N^{gen}_{MC}$ is the total number of generated Monte Carlo (MC) events. The differential cross section, $\sigma^{rad}_{model}(x_j)$, where $j$ refers to the $j$th simulated event in the $i$th $x$ bin, is calculated using the XEMC model (Appendix B of Ref.~\cite{Ye_Thesis}) and includes the radiative effects~\cite{Dasu:1993vk, Moran:2024aou}. These effects account for random energy losses of the incoming and scattered electrons as they pass through the target and detector materials. See Ref.~\cite{Moran:2024aou} for a comparison of these radiative correction procedures to other approaches. The summation runs over all simulated events in the $i$th $x$ bin, applying the same selection criteria as in the experimental data.

The yield-ratio method uses the simulation to account for HRS and detector acceptance, bin centering, and radiative effects. This relies on having a reliable cross-section model, and so after the cross section is extracted using a starting cross-section model~\cite{Fomin:2011ng}, the model is adjusted to match the data so that the cross-section weighting used to evaluate acceptance, radiative corrections, etc. in the simulation is more realistic. This process is repeated until the simulated yield from the updated cross-section model is in good agreement with the data. 

\section{Results}
\subsection{Absolute Cross Sections} \label{sec:cross_section}

\begin{figure}[ht!]
    \centering
        \includegraphics[width=0.46\textwidth]{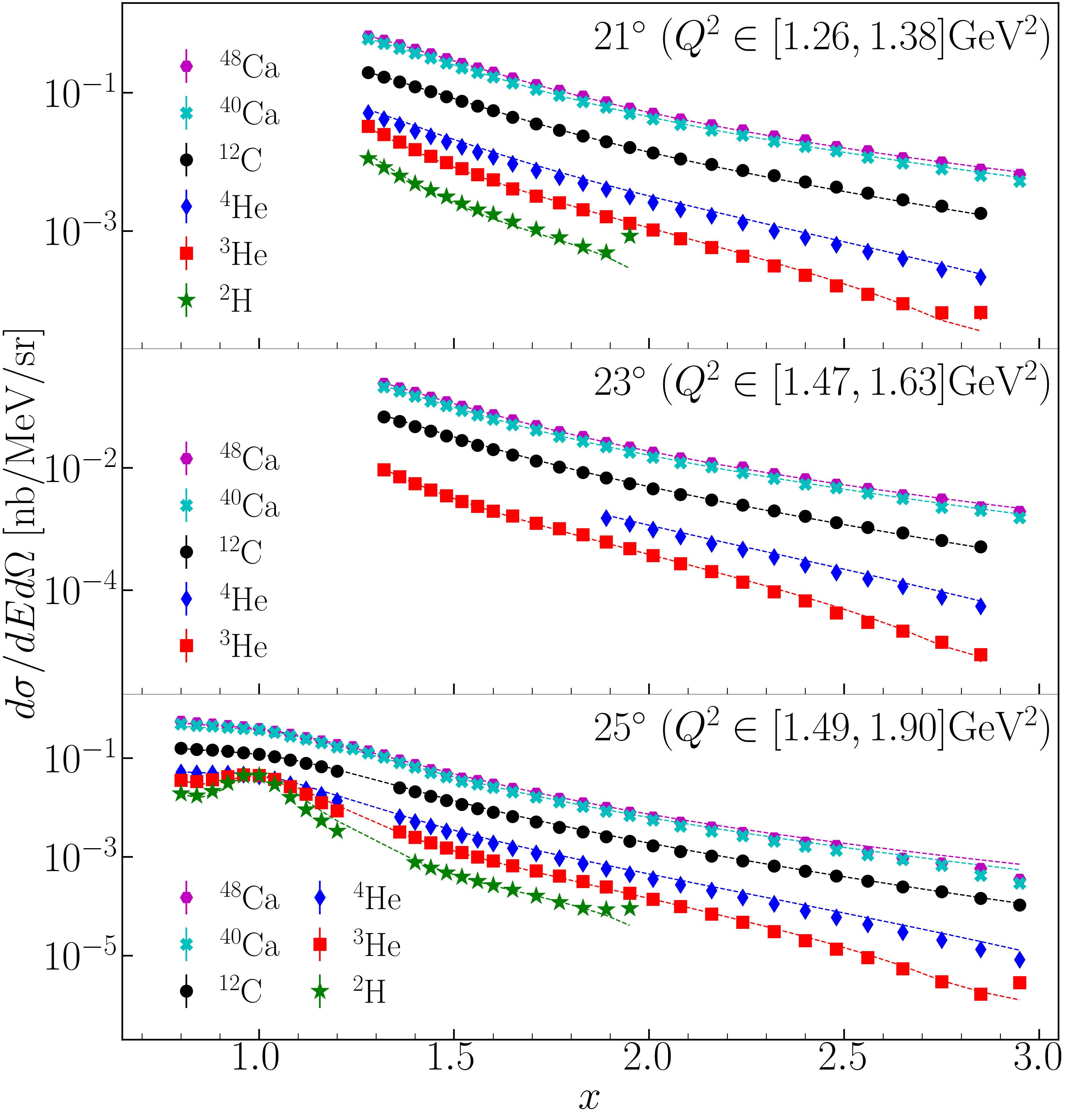}
        \vspace{-0.4cm}
    \caption{Differential cross sections for all nuclear targets for different scattering angles. Scale uncertainties are not shown.}
    \label{fig:XS_21_23_25}
\end{figure}

Figure~\ref{fig:XS_21_23_25} shows the absolute differential cross sections as a function of $x$ for all targets measured in this experiment at scattering angles of 21$^\circ$, 23$^\circ$, and 25$^\circ$. The quasi-elastic cross section for $^2$H should drop rapidly as $x$ approaches the kinematic limit ($x\approx2$) but the total cross section rises due to contributions from e-d elastic scattering. Similarly, $^3$He shows an excess near $x=3$, again indicative of nuclear elastic scattering. The lines are the XEMC cross-section model~\cite{Ye_Thesis}, showing good agreement with the data over the full kinematic range, except for $x \approx$~A, as the model does not include nuclear elastic scattering. The extracted cross sections and fractional uncertainties for all targets at all angles are provided in the supplemental materials~\cite{supplemental}. To explicitly visualize the angular dependence, an additional plot in the supplemental material~\cite{supplemental} compares the absolute differential cross sections of $^3$He and $^{12}$C measured at 21$^\circ$, 23$^\circ$, 25$^\circ$, and 28$^\circ$.

\begin{table}[ht!]
\centering
\begin{tabular}{ | l  | c | c |} 
  \hline
     & $\delta\sigma/\sigma$ (scale) & $\delta\sigma/\sigma$ (p2p)\\ 
   \hline
  
    $\delta_{track}$    & 1\%       &  0.3\% \\   \hline
    $\delta_{PID}$      & 0.3\%     &  0.1\% \\   \hline
    $\delta_{charge}$   & 0.7\%     &  -     \\   \hline
    $\delta_{dt}$       & 0.7\%     &  -     \\   \hline
    $\delta_{target}$   & 5\% (1\%) &  1.5-3.0\% (0\%)  \\   \hline
    $\delta_{kin}$      & 1\%       &  1.5-6\% (1.5-3)\% \\   \hline
    $\delta_{rad}$      &1.5\% (1\%)&  1.0\% (0.5\%) \\   \hline
    $\delta_{acc}$      & 2\%       &  1.5\% \\   \hline
    $\delta_{ext}$      & -         &  2.0-5.0\% \\  \hline
    $\delta_{model}$    & -         &  1.0\% \\   \hline
\end{tabular}
 \caption{Systematic uncertainties for absolute cross sections\label{table_xs}. Where two values are given, the first is for the long targets ($^2$H, $^3$He, and $^4$He), and second (in parentheses) is for the short targets. See text for details of entries that quote a range of values that depend on $x$.}
\end{table}
%



The uncertainties presented in Figures~\ref{fig:XS_21_23_25} include the combined statistical and uncorrelated (point-to-point, p2p) systematic uncertainties. Additionally, each data subset (angle-target combination) has an associated scale uncertainty, which allows for the rescaling of all data points. The scale and p2p uncertainties are summarized in Table~\ref{table_xs}, where the entries correspond to the uncertainty associated with various sources: spectrometer acceptance ($\delta_{acc}$), extended target acceptance ($\delta_{ext}$), tracking efficiency ($\delta_{track}$), PID cuts ($\delta_{PID}$), model dependence of the cross-section extraction ($\delta_{model}$), target thickness ($\delta_{target}$), beam charge measurement ($\delta_{charge}$), electronic and computer dead time ($\delta_{dt}$), and radiative corrections ($\delta_{rad}$). 

Most of these uncertainties are largely independent of kinematics and target, but directly related to the performance of the HRS spectrometer and detectors. Standard HRS analysis procedures were used to study and evaluate the values of $\delta_{track}$, $\delta_{PID}$, $\delta_{charge}$, and $\delta_{dt}$ which all are small. For $\delta_{target}$, the extended targets have larger uncertainties associated with the beam heating effects discussed in the previous section, and an additional p2p uncertainty associated with the combination of a $Z_{react}$-dependent density and uncertainty in the extended-target acceptance. The p2p uncertainty of $\delta_{target}$ for long targets increases linearly from 1.5\% to 3\% as $x$ increases from 1.4 to 2, and continues this linear increase beyond $x=2$. The $\delta_{kin}$ term accounts for the variation of cross section within the uncertainties of the beam and scattered electron energies and angles. These are similar for all targets and all $x$ values, but increase at larger $x$ where the cross section drops rapidly, especially for the $^2$H data approaching $x=2$. The p2p uncertainty of $\delta_{kin}$ for short targets has a linear increase from 1.5\% to 3\% as the value of $x$ increases from 1.5 to 1.8. For long targets, the p2p uncertainty of $\delta_{kin}$ also increases linearly, ranging from 1.5\% to 6.0\%, except for $^2$H whose uncertainty reaches 10\% when $x=1.8$. In both cases, the $\delta_{kin}$ uncertainty continues to increase linearly beyond $x=1.8$ at the rate described above. The extracted cross sections and fractional uncertainties are provided in the supplemental materials~\cite{supplemental}. 

The acceptance of the spectrometers is generally well understood for short targets, in particular because our cuts on the electron momentum and solid angle are tighter than used by many experiments. However, it is not as well known for the 20~cm targets as it is for the thin targets. By examining the variation of the extended-target cross section with the momentum and solid angle cut, we find that going from the nominal HRS acceptance cuts to our tighter cuts systematically increases the cross section by about 2\% up to $x \approx 1.4$ and by roughly 5\% above $x \approx 1.8$. As such, we apply an additional systematic uncertainty on the long target cross sections (and ratio of short targets to long targets) of 2\% for $x<1.4$, 5\% for $x>1.8$, with a linear increase between $x=1.4$ and 1.8. Because this effect is correlated with $x$ and at different angles, we shift the extended-target cross section up by half that amount. This allows for the possibility that the cross section would continue to increase with tighter cuts, but also allows for the possibility that our current cuts are sufficient, with only a modest increase in the total chi-squared value. 

The extended target correction is not relevant for the ratio of short targets, and should cancel almost completely in the ratio of long targets. As expected, we find no systematic trend when we vary the cuts on the target ratios in these cases.  For the A/$^3$He ratios for A$\geq$12, we find a systematic decrease in the ratio, consistent with the increase in the extended target cross section. As such, we apply the same 2-5\% uncertainty to the short-to-long target ratios. 

Imperfections in the model can impact the simulated acceptance, radiative effects, and bin-centering corrections. We estimate the uncertainties of the model dependence ($\delta_{model}$) in the analysis by taking our optimized model and applying an additional factor (e.g. $\sigma_{model}/x$ or $\sigma_{model}/Q^2$) that worsens the $x$ or $Q^2$ dependence of the model compared to the data. These `de-tuned' models were then applied to the same procedure of cross-section extraction as discussed in Section 2.2. Cross sections extracted using these `de-tuned' models give visibly worse agreement with the real results, and the typical 1\% impact on the analysis is taken as a conservative estimate of the model dependence of the analysis.



\subsection{\texorpdfstring{A/$^2$H Cross-Section Ratios}{A/2H Cross-Section Ratios}}

We next form the cross-section ratios of heavier nuclei to $^2$H to investigate the 2N-SRC plateau across different nuclei. To focus on the previously observed 2N-SRC scaling region ($Q^2 \geq 1.4$~GeV$^2$, $x \geq 1.4$)~\cite{Arrington:2022sov, Li:2024rzf}, we excluded the 21$^\circ$ data, as it falls within the low-$Q^2$ range. Since there was no 23$^\circ$ data for $^2$H, the A/$^2$H ratios come entirely from the 25$^\circ$ data. In constructing the target ratios, several systematic uncertainties common to all targets partially or completely cancel out in the A/$^2$H ratios. In particular, scale uncertainties associated with the experimental setup and data analysis -— such as $\delta_{acc}$, $\delta_{track}$, $\delta_{PID}$, and $\delta_{model}$ -— almost entirely cancel in the ratios. A summary of the scale and point-to-point (p2p) uncertainties in the target ratios is provided in Table~\ref{table_ratio}.





\begin{table}[ht!]
\centering
\begin{tabular}{ | c  | c | c |}
\hline
&$\delta R/R$ (scale) & $\delta R/R$ (p2p)\\    \hline
$\delta_{acc} (s/l)$    & 2\%           &  0.5\%        \\   \hline
$\delta_{track}$        & 0.2\%         &  0.2\%        \\  \hline
$\delta_{PID}$          & -             &  0.1\%        \\     \hline
$\delta_{model}$        & -             &  1.0\%        \\ \hline
$\delta_{charge}$       & 1.0\%         &  -            \\  \hline
$\delta_{dt}$           & 0.5\%         &  -            \\   \hline
$\delta_{target}$(s/l)  & 5\%           & 1.5-3.0\%     \\   
$\delta_{target}$(l/l)  & 4\%           & 1.0-2.0\%     \\   
$\delta_{target}$(s/s)  & 1.4\%         & 0\%           \\   \hline
$\delta_{rad}$(s/l, s/s)& 1.5\%         & 1\%           \\   
$\delta_{rad}$(l/l)     & 0.5\%         & 0.3\%         \\  \hline
$\delta_{kin}$($^{3,4}$He/D)& 0.6\%     & 2.0-4.0\%     \\ 
$\delta_{kin}$(A/D)     & 0.6\%         & 1.5-6.0\%     \\ 
$\delta_{kin}$(A/He)    & 0.6\%         & 1.5-3.0\%     \\ 
$\delta_{kin}$(s/s)     & 0.6\%         & 0.7-1.4\%     \\  \hline
\end{tabular}
\caption{Systematic uncertainties for the cross-section ratios. In some cases, the uncertainty depends on the target ratios being taken: s/s, l/l, and s/l refer to the ratio of two short targets, two long targets, or the ratio of a short to long target, respectively, while A/D refer to the ratio of any target to deuterium. The entries that quote a range of values refer to a linear increase in uncertainty with respect to $x$, as in Table~\ref{table_xs}. For $\delta_{kin}$, A/D refers to the solid targets (A$\ge$12).}\label{table_ratio}
\end{table}

While we have thus far used $x$ as a surrogate for the initial struck nucleon’s momentum, the light-cone momentum fraction, $\alpha$, provides a more accurate estimate of the initial momentum~\cite{Frankfurt:1988nt, Frankfurt:1993sp, Arrington:2022sov}.
The light-cone momentum $\alpha$ cannot be reconstructed from inclusive scattering without making an assumption about the unmeasured final state, but $\alpha_{2N}$~\cite{Frankfurt:1993sp} is typically used as the experimental estimate of $\alpha$ in the mean-field and 2N-SRC regimes, based on the assumption that the momentum of the struck nucleon is balanced by one other nucleon:
\begin{equation}
    \alpha_{2 N}=2-\frac{q_{\min }+2 M}{2 M}\left(1+\frac{\sqrt{W^2-4 M^2}}{W}\right)
    \label{eq_alpha_2N}
\end{equation}
where $q_{\min }=\nu-|\vec{q}|$, $W^2=\sqrt{-q^2+4 M v+4 M^2}$.

\begin{figure}[htb!]
    \centering
    \includegraphics[width=0.48\textwidth]{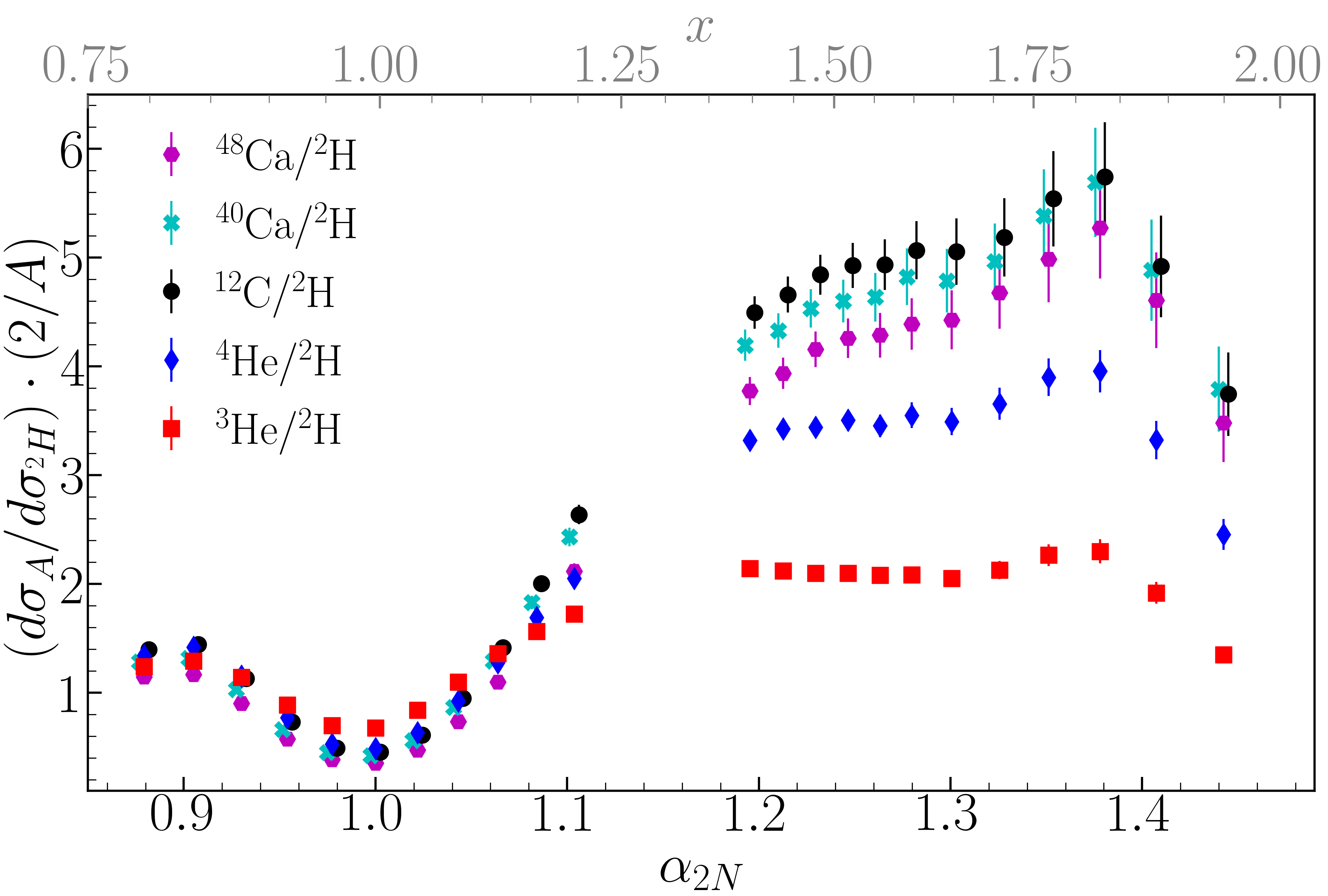}
    \vspace{-0.4cm}
    \caption{The A/$^2$H ratio of the cross section per nucleon as a function of $x$ (top axis label) and $\alpha_{2N}$ (bottom) for the 25$^\circ$ data. Scale uncertainties are not shown. The points for $^{12}$C ($^{40}$Ca) are shifted slightly to the right (left) so they can be more easily distinguished.}
    \label{fig:ratio_D2}
\end{figure}

Fig.~\ref{fig:ratio_D2} shows the A/$^2$H cross-section ratios as functions of $x$ (top) and $\alpha_{2N}$ (bottom), with a gap in coverage for $1.2<x<1.4$ ($1.12 < \alpha_{2N} < 1.18$).
Above $x \approx 1.7$ ($\alpha_{2N} \approx 1.33$), the ratios show a slight increase, reflecting an enhanced cross section for A~$>2$ targets due to smearing from the center-of-mass motion of the 2N-SRC pairs, as seen in previous studies~\cite{Fomin:2011ng}. As $x$ approaches 2, the ratio begins to drop due to the contribution of the deuteron elastic peak. This effect is more pronounced than in higher-$Q^2$ data~\cite{Fomin:2011ng} due to the larger elastic contribution at these $Q^2$ values.

For the light nuclei (A=3,4), a clear plateau is observed for $\alpha_{2N} > 1.18$, with minimal deviations until the effects discussed above begin to modify the ratio above $\alpha_{2N} \approx 1.33$. In contrast, for the heavier nuclei (A$\geq$12), there is an approximate plateau, but the increase at larger $\alpha_{2N}$ values is more pronounced compared to previous measurements. We attribute this difference to variations in the acceptance between the thin ($<$1~cm) foil targets for the heavier nuclei and the 20~cm deuterium cryotarget. Modeling the latter is more challenging in this measurement due to the large, strongly position-dependent target heating effects. We therefore add an additional 1.5\% uncertainty to the $a_2$ values for the carbon and calcium targets to account for the fact that the data do not reproduce the plateau seen in previous measurements. This is based on the cut dependence observed when reducing the $x$ range used in the fit. We take this as a common 1.5\% uncertainty for all three targets.

\begin{table}[ht!]
\centering
\begin{tabular}{ c c c c c c }
    \hline
    Nuclei & $^3$He & $^4$He & $^{12}$C &$^{40}$Ca & $^{48}$Ca\\
    \hline
    $a_2$ 
     & 2.103
     & 3.455
     & 4.785
     & 4.483
     & 4.096 \\
$\delta a_2$ (fit)     & 0.021
     & 0.036
     & 0.073
     & 0.070
     & 0.064  \\     
$\delta a_2$ (scale) & 0.089
     & 0.146
     & 0.274
     & 0.257
     & 0.235 \\
 
    \hline
\end{tabular}
    \vspace{-0.3cm}
\caption{The extracted values of $a_2$ and uncertainties from the present data. $a_2$ were fitted with a flat distribution in the region of $1.18 \le \alpha_{2N} \le 1.33$. Only statistical and point-to-point uncertainties were included in the fit; the scale uncertainty on the cross section ratios is applied directly to $a_2$. For the solid targets (A$\ge$12), there is an additional 1.5\% uncertainty, associated with the model dependence in extracting $a_2$ when the data do not yield a perfect plateau (see text above).}
\label{table_a2_fitting}
\end{table}

\begin{figure}[ht]
\centering
    \includegraphics[width=0.48\textwidth, trim={30 30 5 5}, clip]{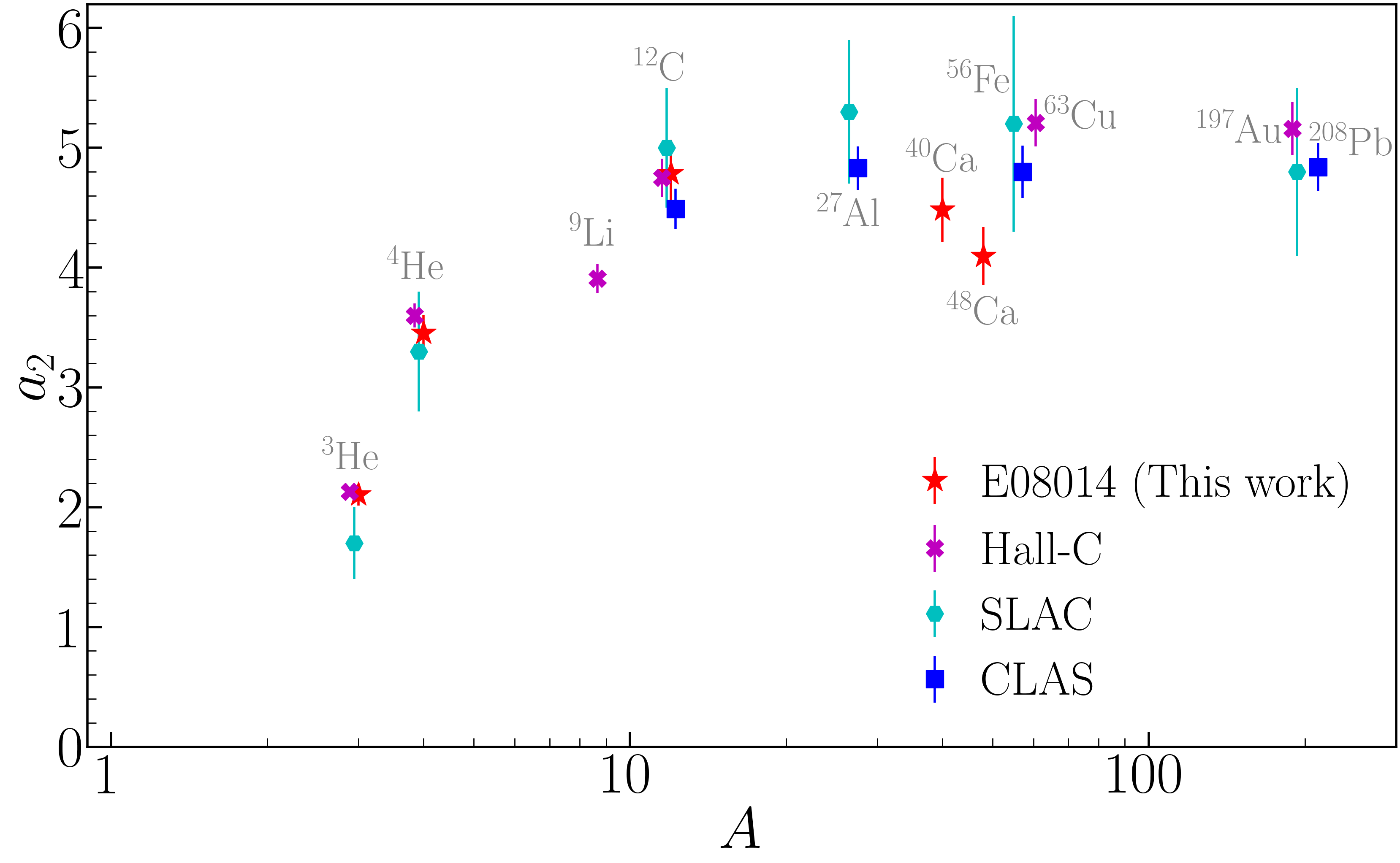}
        \vspace{-0.4cm}
\caption{$a_2$ vs. A where new results (red stars) are from Table ~\ref{table_a2_fitting}, and the $a_2$ values from the previous experiments~\cite{Fomin:2011ng, Frankfurt:1993sp, CLAS:2019vsb} are taken from the updated values in Ref.~\cite{Arrington:2022sov}. Different $a_2$ values for the same $A$ were shifted slightly for better visualization. For the solid targets (A$\ge$12), there is an additional 1.5\% uncertainty (not shown) that affects all three of these targets together, associated with the model dependence in extracting $a_2$ when the data do not yield a perfect plateau (see text above). A table of $a_2$ values from all experiments is included in the supplemental material~\cite{supplemental}.}
    \label{fig:a2_A}
\end{figure}

As discussed in the introduction, the A/$^2$H ratio in the plateau region, denoted as $a_2$(A), represents the relative contribution of 2N-SRCs in nucleus A compared to that in $^2$H. This is dominantly due to the increase of the relative probability for nucleons to be in SRCs in heavier nuclei, but also includes corrections related to binding differences as well as the center-of-mass motion of the 2N-SRC pairs in the A~$>2$ nucleus~\cite{Fomin:2011ng, Weiss:2020bkp}. In addition, any inelastic contribution to the cross section in the SRC regime would modify the cross-section ratio from the pure quasielastic contribution. We do not apply a correction for the small inelastic contribution, as we estimate that this contribution is at the 1\% for $x=1.4$, and the inelastic contributions for the two targets will yield partial cancellation in the ratios, reducing the impact even further. 

Table~\ref{table_a2_fitting} shows the extracted values of $a_2$, obtained by fitting the plateau region for $1.4 \le x \le 1.75$ ($1.18 \le \alpha_{2N} \le 1.33$). As shown in Fig.~\ref{fig:a2_A}, $a_2(^3$He), $a_2(^4$He), and $a_2(^{12}$C) are consistent with previous measurements within their errors. Because the 2N-SRC region does not show as clear of a plateau as previous data for A$\geq$12 as previous measurements did, we apply an additional uncertainty to the extracted $a_2$ values. We look at the variation of the extracted $a_2$ value if we remove the lowest $x$ points, the two highest $x$ points, or all three, and use the scatter as an estimate of the additional uncertainty. This yields an additional uncertainty that is expected to be correlated for the $a_2$ values of these three nuclei.

The values of $a_2(^{40}$Ca) and $a_2(^{48}$Ca) are newly available in this work, and their similar values confirm that the nucleons in the minority (protons) primarily determine the probabilities for forming SRC pairs. For the direct comparison of calcium isotopes, the result of Ref.~\cite{Nguyen:2020wrr} yields the more precise result, yielding $^{48}$Ca/$^{40}$Ca $= 0.971\pm0.012$. As noted above, the inclusion of longer targets for the light nuclei in this work required tighter cuts, reducing the statistical precision compared to Ref.~\cite{Nguyen:2020wrr}. In addition, the previous work performed a more detailed examination of the correlated uncertainties in the calcium target ratio.

\subsection{\texorpdfstring{A/$^3$He Cross-Section Ratios}{A/3He Cross-Section Ratios}}

For three-nucleon SRCs, a different estimate of the light cone momentum, $\alpha$ must be used. We use $\alpha_{3N}$ from ref.~\cite{Sargsian:2019joj}, which is based on the expectation that linear configurations dominate. For the kinematics of this experiment, other configurations are strongly suppressed and due to the increased excitation of the 3N system~\cite{Li:2024rzf}. The expression for $\alpha_{3N}$ used in this paper is included in the supplemental material~\cite{supplemental}.

\begin{figure}[ht!]
    \centering
        \includegraphics[width=0.46\textwidth]{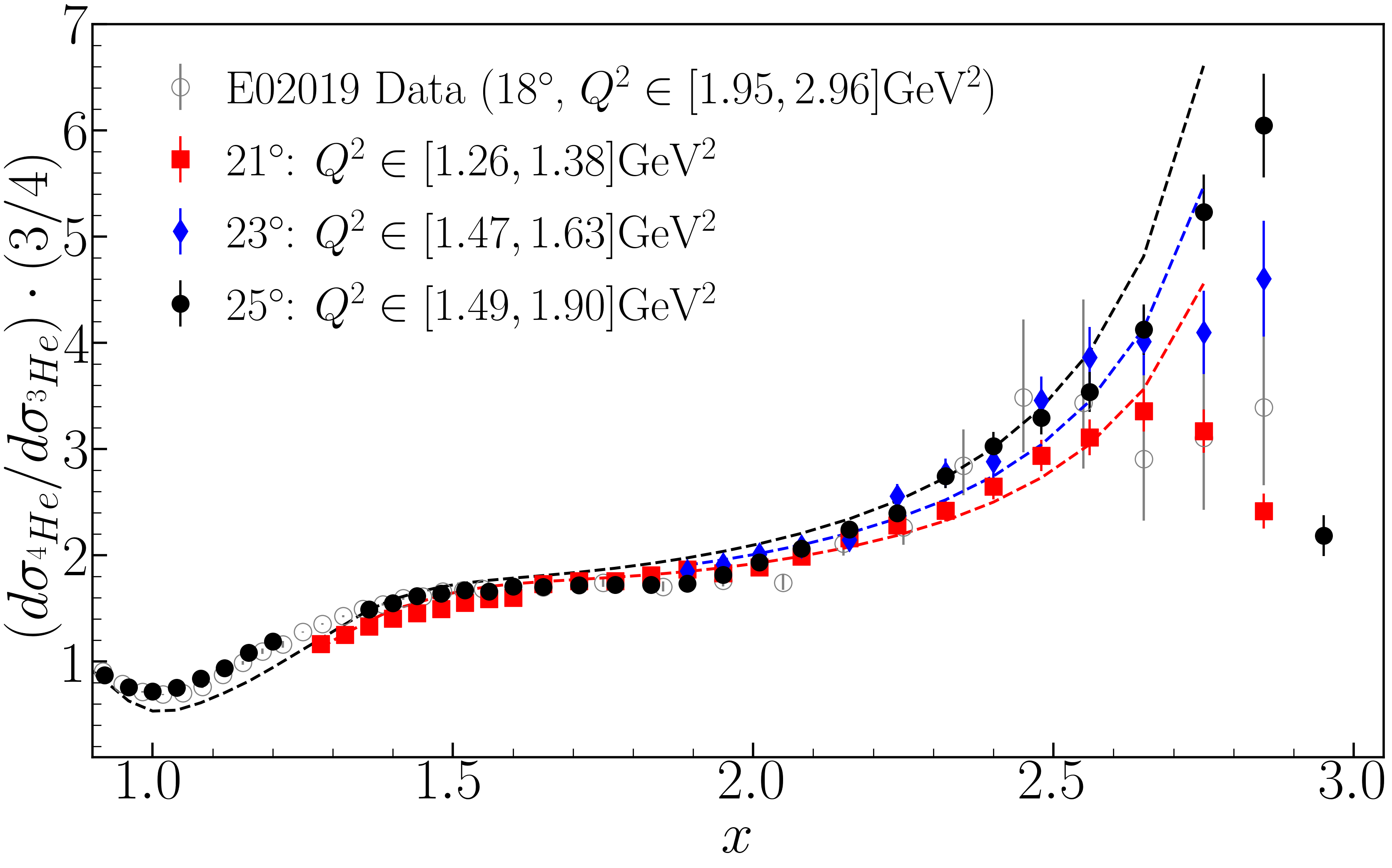}
        \includegraphics[width=0.46\textwidth]{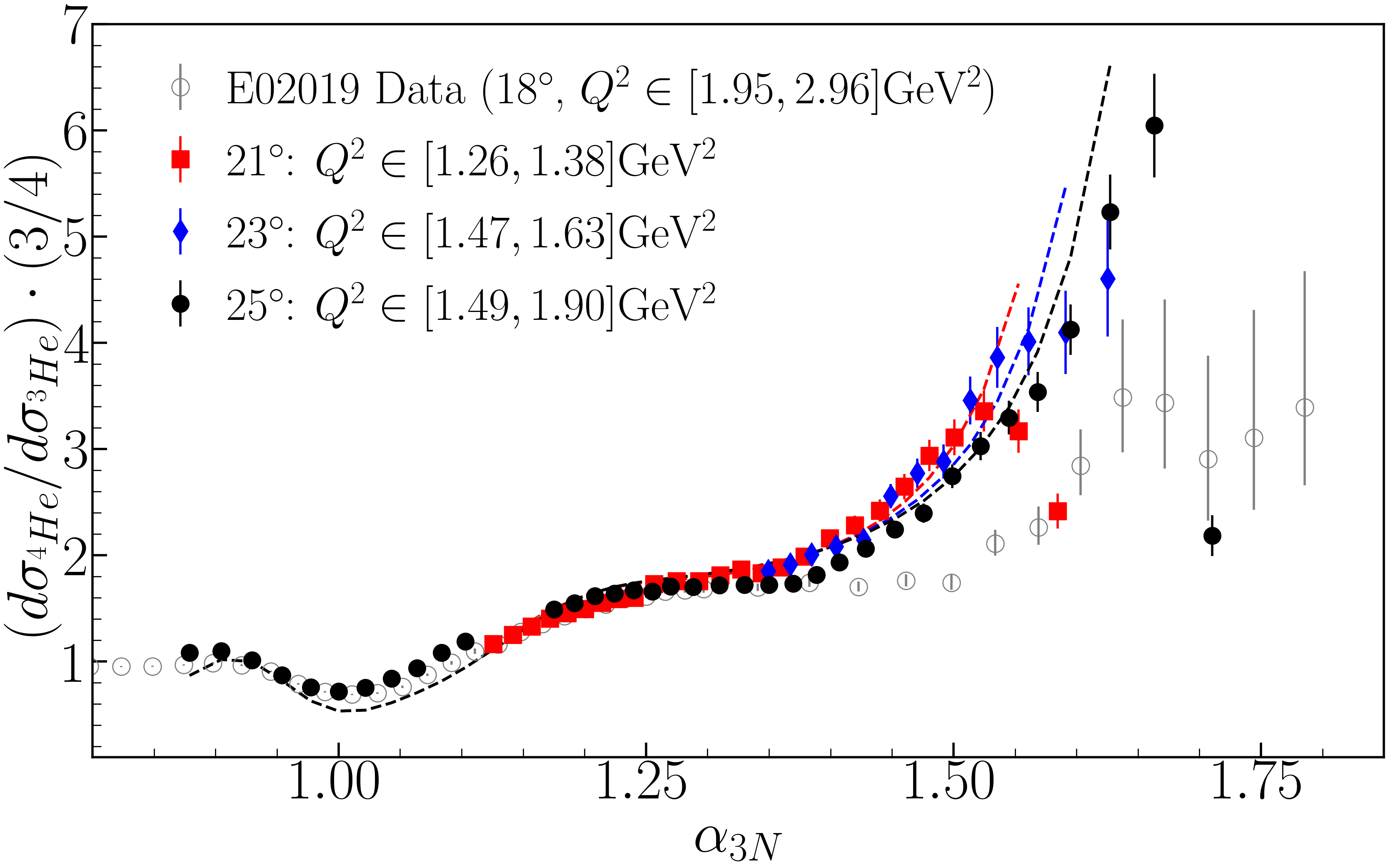}
        \vspace{-0.4cm}
    \caption{The $^4$He/$^3$He ratio of the cross section per nucleon as a function of $x$ (top) $\alpha_{3N}$ (bottom) for all data sets from this work and from the lowest $Q^2$ data set of Ref.~\cite{Fomin:2011ng}. Scale uncertainties are not shown. Lines are the ratios of cross sections calculated using the XEMC model, included to allow for comparison of the $Q^2$ dependence of the $y$-scaling model.}
    \label{fig:E02019_He4_He3}
\end{figure}

\begin{figure}[ht!]
    \centering
        \includegraphics[width=0.46\textwidth]{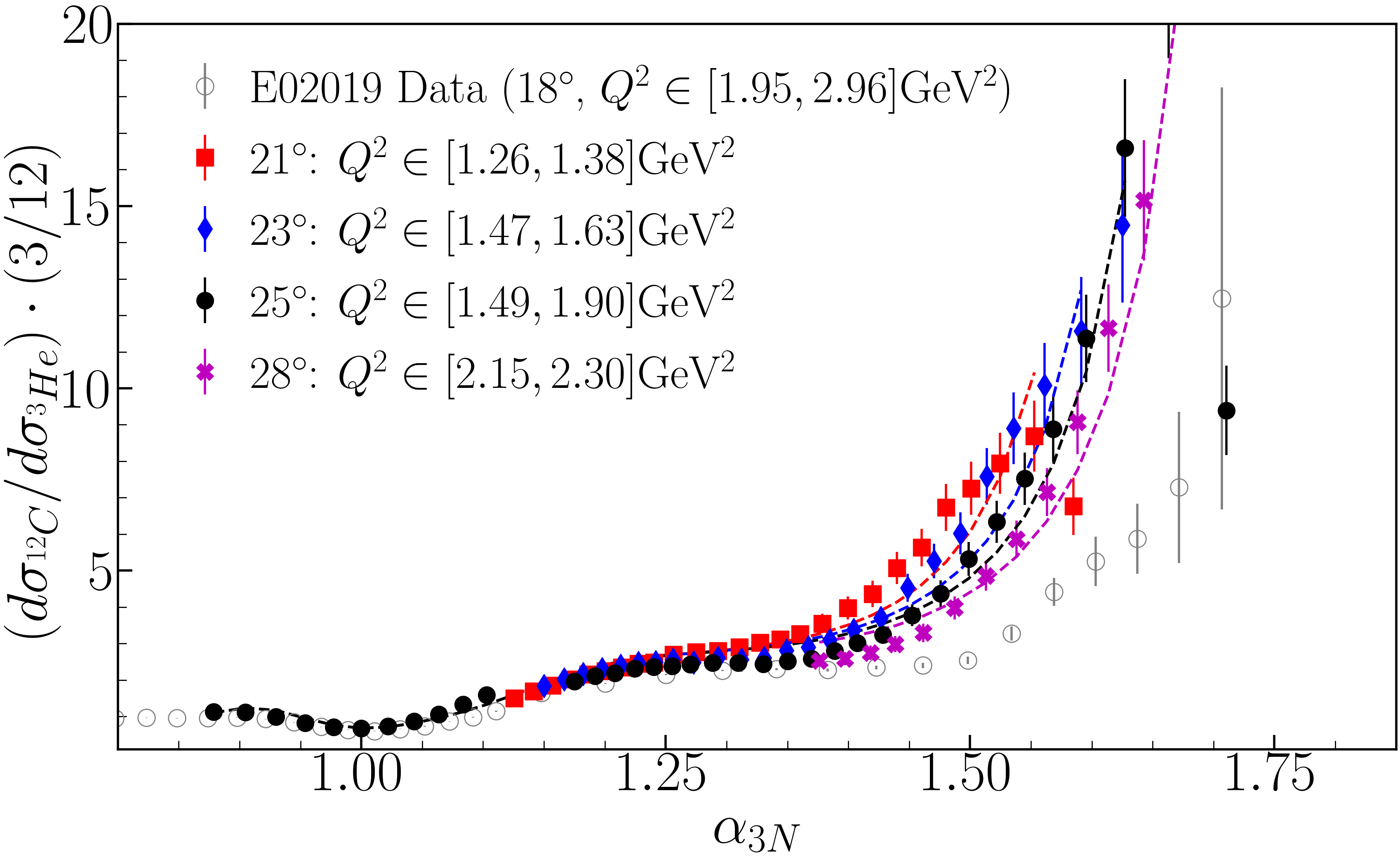}
    \vspace{-0.4cm}    
    \caption{The $^{12}C$/$^3$He ratio of the cross section per nucleon as a function of $x$ (top) $\alpha_{3N}$ (bottom) for all angles. Data are from this work and Ref.~\cite{Fomin:2011ng}. Scale uncertainties are not shown. Lines are the ratios of cross sections calculated using the XEMC model.}
    \label{fig:E02019_C12_He3}
\end{figure}

Figure~\ref{fig:E02019_He4_He3} shows the $^4$He/$^3$He ratios at all scattering angles as a function of $x$ and $\alpha_{3N}$, including the E02-019 results at higher $Q^2$ for comparison~\cite{Fomin:2011ng}. As reported in Ref.~\cite{Ye:2017ivm} which shows the $^4$He/$^3$He ratio with combined 23$^{\circ}$ and 25$^{\circ}$ data, there is no indication of the onset of a 3N-SRC plateau. Instead, a rapid increase is observed for $x>2$ ($\alpha_{3N}>1.41$) as $Q^2$ increases, moving the ratios away from the existing higher-$Q^2$ data~\cite{Fomin:2011ng}. This was unexpected as nearly all of the scaling violating effects - FSI, CM motion, and binding of the 2N-SRC - have been seen to decrease with $Q^2$ in detailed scaling studies for 2N-SRCs~\cite{Frankfurt:1993sp, CLAS:2003eih, Li:2024rzf}. In addition, the data are moving away from the predicted plateau expected if 3N-SRCs dominate.

However, the situation is qualitatively different when the ratios are plotted against $\alpha_{3N}$. At these $Q^2$ values, the difference between $x$ and $\alpha_{3N}$ is strongly $Q^2$ dependent, and with the rapid rise of the $^4$He/$^3$He ratios with increasing $x$, even a modest shift between data sets going from $x$ to $\alpha_{3N}$ has a significant impact on the ratios. When examined as a function of $\alpha_{3N}$, the ratios in the 3N-SRC region decrease as $Q^2$ increases, consistent with the expectation of a plateau at large $Q^2$ values and bringing the data into better agreement with the higher-$Q^2$ measurements (except for $x \approx 3$, where nuclear elastic contributions enhance the denominator). This can be seen even more clearly in the $^{12}$C/$^3$He ratios shown in Fig.~\ref{fig:E02019_C12_He3}, where there is an additional data set at higher $Q^2$. This new examination of the $Q^2$ dependence for the large-$\alpha_{3N}$ data supports the argument that larger $Q^2$ values are needed to try and isolate 3N-SRCs~\cite{Fomin:2017ydn, Day:2018nja, Sargsian:2019joj, Fomin:2023gdz}.

A plot showing the A/$^3$He ratios for all targets is included in the supplemental materials~\cite{supplemental}, but the solid targets are all very similar to the carbon ratios shown in Fig.~\ref{fig:E02019_C12_He3}. The cross-section ratios of A to $^2$H ($^3$He) as functions of $x$ and $\alpha_{2N}$ ($\alpha_{3N}$) are also given in the supplemental materials~\cite{supplemental} for all of the targets measured in this experiment.

\section{Conclusion}

In this work, we presented the complete set of cross sections and ratios from inclusive quasi-elastic scattering data obtained in Jefferson Lab experiment E08-014, examining nuclei ranging from $^2$H to $^{48}$Ca. We find a plateau in the A/$^2$H ratios for $x>1.4$ ($\alpha_{2N}>1.18$), consistent with previous data and predictions based on the dominance of 2N-SRCs. 
No 3N-SRC plateau was observed in the $^4$He/$^3$He and $^{12}$C/$^3$He ratios at $x>2$, where the coverage in $\alpha_{2N}$ was limited to $\alpha_{2N}\ltorder 1.6$, slightly below the region where recent predictions suggest 3N-SRCs may dominate~\cite{Sargsian:2019joj, Day:2018nja}. While the $^4$He/$^3$He ratios taken vs $x$ appear to be moving away from the predicted plateau as $Q^2$ increases~\cite{Ye:2018jth}, the A/$^3$He ratios decrease with increasing $Q^2$, especially for $^{12}$C where the $Q^2$ range is larger, consistent with an approach to the predicted 3N-SRC dominance. 
The fact that the data are approaching the high-$Q^2$, but limited precision Hall C data~\cite{Fomin:2011ng} suggests that high-precision measurements at larger $Q^2$ values, such as those taken by JLab experiment E12-06-105~\cite{e12-06-105} may be able to isolate the presence of 3N-SRCs. 




\section*{Acknowledgments}
We gratefully acknowledge the exceptional support from the Hall A technical staff and the JLab target group. This work was partially supported by the Department of Energy’s Office of Science, Office of Nuclear Physics, under Contracts No. DE-AC05-06OR23177, DE-AC02-05CH11231, DE-AC02-06CH11357, DE-FG02-96ER40950, and DE-SC0013615 as well as by the National Science Foundation. Additional support under DOE Contract No. DE-AC05-06OR23177 was provided through JSA, LLC, which operates JLab. Y.P. Zhang and Z.H. Ye acknowledge support from the National Science Foundation of China under contract 12275148 and 12361141822 and the Tsinghua University Initiative Scientific Research Program. A.~Shahinyan acknowledges support from Grant 21AG-1C085 by the Science Committee of the Republic of Armenia. The $^{48}$Ca isotope used in this research was supplied by the Isotope Program within the Office of Nuclear Physics in the Department of Energy’s Office of Science. Finally, we remember and honor Patricia Solvignon, who was instrumental in the development of Experiment E08-014; her passing is deeply felt by our community.

\section*{Supplementray Materials}
Supplementary material, including detailed kinematic settings, target density calibrations, additional plots and tables of all cross-sections and ratios, is available online at Ref.~\cite{suppl_plb}. The cross-section results in data format are available on request.

\bibliographystyle{elsarticle-num}

\bibliography{main}

@misc{supplemental,
    note  = "See Supplemental Material at [URL will be inserted by publisher] for detailed expressions and tables of cross section results.",
year="2025"
}

@misc{E12-06-105,
   author = "Arrington, J and Day, D and Fomin,N and Solvignon-Slifer, P",
   title = "{E12-06-105: Inclusive Scattering from Nuclei at $x > 1$ in the quasielastic and deeply inelastic regimes}",
   url = "www.jlab.org/exp_prog/proposals/06/PR12-06-105.pdf",
   year = "2006"
}

@misc{E08-014,
   author = "Solvignon-Slifer, P and Arrington, J and Day, D and Higinbotham,D",
   title = "{E08-014: Three-nucleon short range correlation studies in inclusive scattering for $0.8 < Q^2 < 2.8$~GeV$^2$.}",
   url = "http://www.jlab.org/exp_prog/proposals/08/PR-08-014.pdf",
   year = "2008"
}

@article{CLAS:2010yvl,
    author = "Baghdasaryan, H. and others",
    collaboration = "CLAS",
    title = "{Tensor Correlations Measured in $^3He(e,e'pp)n$}",
    eprint = "1008.3100",
    archivePrefix = "arXiv",
    primaryClass = "nucl-ex",
    reportNumber = "JLAB-PHY-10-1237",
    doi = "10.1103/PhysRevLett.105.222501",
    journal = "Phys. Rev. Lett.",
    volume = "105",
    pages = "222501",
    year = "2010"
}

@article{Lapikas:1999ss,
    author = "Lapikas, Louk and van der Steenhoven, G. and Frankfurt, L. and Strikman, M. and Zhalov, M.",
    title = "{The Transparency of C-12 for protons}",
    eprint = "nucl-ex/9905009",
    archivePrefix = "arXiv",
    doi = "10.1103/PhysRevC.61.064325",
    journal = "Phys. Rev. C",
    volume = "61",
    pages = "064325",
    year = "2000"
}

@article{CLAS:2018xvc,
    author = "Duer, M. and others",
    collaboration = "CLAS",
    title = "{Direct Observation of Proton-Neutron Short-Range Correlation Dominance in Heavy Nuclei}",
    eprint = "1810.05343",
    archivePrefix = "arXiv",
    primaryClass = "nucl-ex",
    doi = "10.1103/PhysRevLett.122.172502",
    journal = "Phys. Rev. Lett.",
    volume = "122",
    pages = "172502",
    year = "2019"
}

@article{CLAS:2020mom,
    author = "Schmidt, A. and others",
    collaboration = "CLAS",
    title = "{Probing the core of the strong nuclear interaction}",
    eprint = "2004.11221",
    archivePrefix = "arXiv",
    primaryClass = "nucl-ex",
    doi = "10.1038/s41586-020-2021-6",
    journal = "Nature",
    volume = "578",
    pages = "540",
    year = "2020"
}

@article{Weiss:2020bkp,
    author = "Weiss, R. and Denniston, A. W. and Pybus, J. R. and Hen, O. and Piasetzky, E. and others",
    title = "{Extracting the number of short-range correlated nucleon pairs from inclusive electron scattering data}",
    eprint = "2005.01621",
    archivePrefix = "arXiv",
    primaryClass = "nucl-th",
    doi = "10.1103/PhysRevC.103.L031301",
    journal = "Phys. Rev. C",
    volume = "103",
    pages = "L031301",
    year = "2021"
}

@article{Sargsian:2002wc,
    author = "Sargsian, M. M. and others",
    title = "{Hadrons in the nuclear medium}",
    eprint = "nucl-th/0210025",
    archivePrefix = "arXiv",
    reportNumber = "NT-UW-02-020, JLAB-PHY-02-48",
    doi = "10.1088/0954-3899/29/3/201",
    journal = "J. Phys. G",
    volume = "29",
    pages = "R1",
    year = "2003"
}

@article{Arrington:2011xs,
    author = "Arrington, J. and Higinbotham, D. W. and Rosner, G. and Sargsian, M.",
    title = "{Hard probes of short-range nucleon-nucleon correlations}",
    eprint = "1104.1196",
    archivePrefix = "arXiv",
    primaryClass = "nucl-ex",
    reportNumber = "PHY-12946-ME-2011, JLAB-PHY-11-1329",
    doi = "10.1016/j.ppnp.2012.04.002",
    journal = "Prog. Part. Nucl. Phys.",
    volume = "67",
    pages = "898",
    year = "2012"
}

@article{Frankfurt:1988nt,
    author = "Frankfurt, L. L. and Strikman, M. I.",
    title = "{Hard Nuclear Processes and Microscopic Nuclear Structure}",
    doi = "10.1016/0370-1573(88)90179-2",
    journal = "Phys. Rept.",
    volume = "160",
    pages = "235",
    year = "1988"
}

@article{Frankfurt:1993sp,
    author = "Frankfurt, L. L. and Strikman, M. I. and Day, D. B. and Sargsian, M.",
    title = "{Evidence for short range correlations from high $Q^2$ (e, e-prime) reactions}",
    doi = "10.1103/PhysRevC.48.2451",
    journal = "Phys. Rev. C",
    volume = "48",
    pages = "2451",
    year = "1993"
}

@article{Higinbotham:2009zz,
    author = "Higinbotham, Douglas and Piasetzky, Eli and Strikman, Mark",
    title = "{Protons and neutrons cosy up in nuclei and neutron stars}",
    reportNumber = "JLAB-PHY-09-936",
    journal = "CERN Cour.",
    volume = "49N1",
    pages = "22--24",
    year = "2009"
}

@article{CLAS:2005ola,
    author = "Egiyan, K. S. and others",
    collaboration = "CLAS",
    title = "{Measurement of 2- and 3-nucleon short range correlation probabilities in nuclei}",
    eprint = "nucl-ex/0508026",
    archivePrefix = "arXiv",
    reportNumber = "JLAB-PHY-05-285",
    doi = "10.1103/PhysRevLett.96.082501",
    journal = "Phys. Rev. Lett.",
    volume = "96",
    pages = "082501",
    year = "2006"
}

@article{Fomin:2017ydn,
    author = "Fomin, Nadia and Higinbotham, Douglas and Sargsian, Misak and Solvignon, Patricia",
    title = "{New Results on Short-Range Correlations in Nuclei}",
    eprint = "1708.08581",
    archivePrefix = "arXiv",
    primaryClass = "nucl-th",
    reportNumber = "NUPAR-08-2017-1",
    doi = "10.1146/annurev-nucl-102115-044939",
    journal = "Ann. Rev. Nucl. Part. Sci.",
    volume = "67",
    pages = "129",
    year = "2017"
}

@article{Hen:2016kwk,
    author = "Hen, O. and Miller, G. A. and Piasetzky, E. and Weinstein, L. B.",
    title = "{Nucleon-Nucleon Correlations, Short-lived Excitations, and the Quarks Within}",
    eprint = "1611.09748",
    archivePrefix = "arXiv",
    primaryClass = "nucl-ex",
    doi = "10.1103/RevModPhys.89.045002",
    journal = "Rev. Mod. Phys.",
    volume = "89",
    pages = "045002",
    year = "2017"
}

@article{Sargsian:2019joj,
    author = "Sargsian, Misak M. and Day, Donal B. and Frankfurt, Leonid L. and Strikman, Mark I.",
    title = "{Searching for three-nucleon short-range correlations}",
    eprint = "1910.14663",
    archivePrefix = "arXiv",
    primaryClass = "nucl-th",
    reportNumber = "NUPAR-10-2019-01",
    doi = "10.1103/PhysRevC.100.044320",
    journal = "Phys. Rev. C",
    volume = "100",
    pages = "044320",
    year = "2019"
}

@article{Fomin:2011ng,
    author = "Fomin, N. and others",
    title = "{New measurements of high-momentum nucleons and short-range structures in nuclei}",
    eprint = "1107.3583",
    archivePrefix = "arXiv",
    primaryClass = "nucl-ex",
    reportNumber = "JLAB-PHY-11-1408",
    doi = "10.1103/PhysRevLett.108.092502",
    journal = "Phys. Rev. Lett.",
    volume = "108",
    pages = "092502",
    year = "2012"
}

@article{CLAS:2003eih,
    author = "Egiyan, K. S. and others",
    collaboration = "CLAS",
    title = "{Observation of nuclear scaling in the A(e, e$^\prime$) reaction at $x_B > 1$}",
    eprint = "nucl-ex/0301008",
    archivePrefix = "arXiv",
    reportNumber = "JLAB-PHY-03-48",
    doi = "10.1103/PhysRevC.68.014313",
    journal = "Phys. Rev. C",
    volume = "68",
    pages = "014313",
    year = "2003"
}

@article{Nguyen:2020wrr,
    author = "Nguyen, D. and others",
    title = "{Novel observation of isospin structure of short-range correlations in calcium isotopes}",
    reportNumber = "JLAB-PHY-20-3177, DOE/OR/23177-4956",
    doi = "10.1103/PhysRevC.102.064004",
    journal = "Phys. Rev. C",
    volume = "102",
    pages = "064004",
    year = "2020"
}

@article{Ye:2017ivm,
    author = "Ye, Z. and others",
    title = "{Search for three-nucleon short-range correlations in light nuclei}",
    reportNumber = "JLAB-PHY-18-2639",
    doi = "10.1103/PhysRevC.97.065204",
    journal = "Phys. Rev. C",
    volume = "97",
    pages = "065204",
    year = "2018"
}

@article{Arrington:1998ps,
    author = "Arrington, J. and others",
    title = "{Inclusive electron - nucleus scattering at large momentum transfer}",
    eprint = "nucl-ex/9811008",
    archivePrefix = "arXiv",
    reportNumber = "OAP-747",
    doi = "10.1103/PhysRevLett.82.2056",
    journal = "Phys. Rev. Lett.",
    volume = "82",
    pages = "2056",
    year = "1999"
}

@article{Arrington:2012ax,
    author = "Arrington, John and Daniel, Aji and Day, Donal and Fomin, Nadia and Gaskell, Dave and Solvignon, Patricia",
    title = "{A detailed study of the nuclear dependence of the EMC effect and short-range correlations}",
    eprint = "1206.6343",
    archivePrefix = "arXiv",
    primaryClass = "nucl-ex",
    reportNumber = "JLAB-PHY-12-1587",
    doi = "10.1103/PhysRevC.86.065204",
    journal = "Phys. Rev. C",
    volume = "86",
    pages = "065204",
    year = "2012"
}

@article{Ye:2018jth,
    author = "Ye, Zhihong and Arrington, John",
    title = "{Inclusive Studies of Short-Range Correlations: Overview and New Results, Proceedings of the 13th Conference on the Intersections of Particle and Nuclear Physics}",
    journal = "arXiv:1810.03667",
    year = "2018"
}

@article{CLAS:2019vsb,
    author = "Schmookler, B. and others",
    collaboration = "CLAS",
    title = "{Modified structure of protons and neutrons in correlated pairs}",
    eprint = "2004.12065",
    archivePrefix = "arXiv",
    primaryClass = "nucl-ex",
    doi = "10.1038/s41586-019-0925-9",
    journal = "Nature",
    volume = "566",
    pages = "354",
    year = "2019"
}

@article{Subedi:2008zz,
    author = "Subedi, R. and others",
    title = "{Probing Cold Dense Nuclear Matter}",
    eprint = "0908.1514",
    archivePrefix = "arXiv",
    primaryClass = "nucl-ex",
    reportNumber = "JLAB-PHY-08-828",
    doi = "10.1126/science.1156675",
    journal = "Science",
    volume = "320",
    pages = "1476",
    year = "2008"
}

@article{Korover:2014wqo,
    author = "Korover, I. and others",
    title = "{Probing the Repulsive Core of the Nucleon-Nucleon Interaction via the $^4$He(e,e$\prime$pN) Triple-Coincidence Reaction}",
    eprint = "1401.6138",
    archivePrefix = "arXiv",
    primaryClass = "nucl-ex",
    reportNumber = "JLAB-PHY-14-1862",
    doi = "10.1103/PhysRevLett.113.022501",
    journal = "Phys. Rev. Lett.",
    volume = "113",
    pages = "022501",
    year = "2014"}

@phdthesis{Ye_Thesis,
    author = "Ye, Zhihong",
    title = "{Short Range Correlations in Nuclei at Large xbj through Inclusive Quasi-Elastic Electron Scattering}",
    eprint = "1408.5861",
    archivePrefix = "arXiv",
    primaryClass = "nucl-ex",
    reportNumber = "JLAB-PHY-13-1830, DOE/OR/23177-2974",
    doi = "10.18130/V3FJ7H",
    school = "Physics - Graduate School of Arts and Sciences, University of Virginia",
    year = "2013"
}

@article{Tang:2002ww,
      title          = "{n-p short range correlations from (p,2p + n)
                        measurements}",
  author = {Tang, A. and Watson, J. W. and Aclander, J. and Alster, J. and Asryan, G. and others},
  journal = {Phys. Rev. Lett.},
  volume = {90},
  issue = {4},
  pages = {042301},
  numpages = {4},
  year = {2003},
  publisher = {American Physical Society},
  doi = {10.1103/PhysRevLett.90.042301},
      eprint         = "nucl-ex/0206003",
      archivePrefix  = "arXiv",
      primaryClass   = "nucl-ex",
      reportNumber   = "KSU-CNR-202-07",
      SLACcitation   = "%%CITATION = NUCL-EX/0206003;%%"
}

@article{Shneor:2007lly,
    author = "Shneor, R. and others",
    title = "{Investigation of proton-proton short-range correlations via the C-12(e, e$^\prime$ pp) reaction}",
    eprint = "nucl-ex/0703023",
    archivePrefix = "arXiv",
    reportNumber = "JLAB-PHY-07-624",
    doi = "10.1103/PhysRevLett.99.072501",
    journal = "Phys. Rev. Lett.",
    volume = "99",
    pages = "072501",
    year = "2007"
}

@article{Schiavilla:2006xx,
    author = "Schiavilla, R. and Wiringa, Robert B. and Pieper, Steven C. and Carlson, J.",
    title = "{Tensor Forces and the Ground-State Structure of Nuclei}",
    eprint = "nucl-th/0611037",
    archivePrefix = "arXiv",
    reportNumber = "JLAB-THY-06-562",
    doi = "10.1103/PhysRevLett.98.132501",
    journal = "Phys. Rev. Lett.",
    volume = "98",
    pages = "132501",
    year = "2007"
}

@article{Alvioli:2007zz,
    author = "Alvioli, M. and Ciofi degli Atti, C. and Morita, H.",
    title = "{Proton-neutron and proton-proton correlations in medium-weight nuclei and the role of the tensor force}",
    doi = "10.1103/PhysRevLett.100.162503",
    journal = "Phys. Rev. Lett.",
    volume = "100",
    pages = "162503",
    year = "2008"
}

@article{Li:2022fhh,
    author = "Li, S. and others",
    title = "{Revealing the short-range structure of the mirror nuclei $^{3}$H and $^{3}$He}",
    doi = "10.1038/s41586-022-05007-2",
    journal = "Nature",
    volume = "609",
    pages = "41",
    year = "2022"
}

@article{Li:2024rzf,
    author = "Li, S. and others",
    title = "{Inclusive studies of two- and three-nucleon short-range correlations in 3H and 3He}",
    eprint = "2404.16235",
    archivePrefix = "arXiv",
    primaryClass = "nucl-ex",
    reportNumber = "JLAB-PHY-25-4410",
    doi = "10.1016/j.physletb.2025.139734",
    journal = "Phys. Lett. B",
    volume = "868",
    pages = "139734",
    year = "2025"
}

@article{Fomin:2023gdz,
    author = "Fomin, Nadia and Arrington, John and Li, Shujie",
    title = "{Searching for three-nucleon short-range correlations}",
    eprint = "2309.03963",
    archivePrefix = "arXiv",
    primaryClass = "nucl-ex",
    doi = "10.1140/epja/s10050-023-01112-6",
    journal = "Eur. Phys. J. A",
    volume = "59",
    pages = "205",
    year = "2023"
}

@article{Arrington:2022sov,
    author = "Arrington, John and Fomin, Nadia and Schmidt, Axel",
    title = "{Progress in understanding short-range structure in nuclei: an experimental perspective}",
    eprint = "2203.02608",
    archivePrefix = "arXiv",
    primaryClass = "nucl-ex",
    doi = "10.1146/annurev-nucl-102020-022253",
    journal = "Ann. Rev. Nucl. Part. Sci.",
    volume = "72",
    pages = "307",
    year = "2022"
}

@article{Alcorn:2004sb,
    author = "Alcorn, J. and others",
    title = "{Basic Instrumentation for Hall A at Jefferson Lab}",
    reportNumber = "JLAB-PHY-03-129",
    doi = "10.1016/j.nima.2003.11.415",
    journal = "Nucl. Instrum. Meth. A",
    volume = "522",
    pages = "294",
    year = "2004"
}

@article{Dasu:1993vk,
    author = "Dasu, S. and others",
    title = "{Measurement of kinematic and nuclear dependence of R = $\sigma_L$ / $\sigma_T$ in deep inelastic electron scattering}",
    reportNumber = "SLAC-PUB-5814, UR-1304, ER-40685-753",
    doi = "10.1103/PhysRevD.49.5641",
    journal = "Phys. Rev. D",
    volume = "49",
    pages = "5641",
    year = "1994"
}

@article{CiofidegliAtti:2015lcu,
    author = "Ciofi degli Atti, Claudio",
    title = "{In-medium short-range dynamics of nucleons: Recent theoretical and experimental advances}",
    doi = "10.1016/j.physrep.2015.06.002",
    journal = "Phys. Rept.",
    volume = "590",
    pages = "1--85",
    year = "2015"
}

@article{Day:2018nja,
    author = "Day, Donal B. and Frankfurt, Leonid L. and Sargsian, Misak M. and Strikman, Mark I.",
    title = "{Toward observation of three-nucleon short-range correlations in high-Q2~A(e,e$^\prime$)X reactions}",
    eprint = "1803.07629",
    archivePrefix = "arXiv",
    primaryClass = "nucl-th",
    reportNumber = "NuPar-03-2018-01/06-2022-01, NuPar-03-2018-01, NUPAR-03-2018-01",
    doi = "10.1103/PhysRevC.107.014319",
    journal = "Phys. Rev. C",
    volume = "107",
    pages = "014319",
    year = "2023"
}

@article{Moran:2024aou,
    author = "Moran, S. and Arratia, M. and Arrington, J. and Gaskell, D. and Schmookler, B.",
    title = "{Significance of radiative corrections on measurements of the EMC effect}",
    doi = "10.1103/PhysRevC.110.025202",
    journal = "Phys. Rev. C",
    volume = "110",
    pages = "025202",
    year = "2024"
}

@article{Dalal:2022zkg,
    author = "Dalal, Ranjeet and MacGregor, I. J. Douglas",
    title = "{Nucleon-nucleon correlations inside atomic nuclei: synergies, observations and theoretical models}",
    doi = "10.1088/1361-6633/ad27dd",
    journal = "Rept. Prog. Phys.",
    volume = "87",
    pages = "034301",
    year = "2024"
}

@misc{suppl_plb,
   author = "{Y.P. Zhang, et. al.}",
   title = "{Supplementary materials for the published version of this article on Phys. Lett. B}",
   doi = "https://doi.org/10.1016/j.physletb.2025.140087",
   url = "https://ars.els-cdn.com/content/image/1-s2.0-S0370269325008457-mmc1.pdf",
   year = "2025"
}

\end{document}